**RESEARCH ARTICLE** OPEN ACCESS

# 5G Network Slicing as a Service Enabler for the Automotive Sector

David Candal-Ventureira[1] | José Manuel Rúa-Estévez[1] | Pablo Fondo-Ferreiro[1] | Felipe Gil-Castiñeira[1] | Antonio Fernández-Barciela[2] | Francisco Javier González-Castaño[1] | Emilio Diéguez-Pazo[2] | Luis Fernández-Ferreira[2]

[1]Information Technologies Group, atlanTTic, University of Vigo, Vigo, Spain | [2]Telecommunication and Connectivity, Stellantis, Vigo, Spain

**Correspondence:** Pablo Fondo-Ferreiro (pfondo@gti.uvigo.es)

**Received:** 10 May 2024 | **Revised:** 10 September 2024 | **Accepted:** 29 September 2024

**Funding:** This work was supported by Xunta de Galicia (Spain), ED431C 2022/04, ED481B-2022-019, IN854A 2020/01, IN854A 2023/01. Ministry of Science, Innovation and Universities (MICIU), State Investigation Agency (AEI), Spain, and the European Social Fund (FSE+), PDC2021-121335-C21, PID2020-116329GB-C21, PRE2021-098290, Universidade de Vigo/CISUG for open access.

**Keywords:** 5G | automotive | factory | network slicing | V2X

**ABSTRACT**

Network slicing, a key technology introduced in 5G standards, enables mobile networks to simultaneously support a wide range of heterogeneous use cases with diverse quality of service (QoS) requirements. This work discusses the potential benefits of network slicing for the automotive sector, encompassing manufacturing processes and vehicular communications. The review of the state of the art reveals a clear gap regarding the application of network slicing from the perspective of industrial verticals such as automotive use cases and their specific requirements. Departing from this observation, we first identify limitations of previous cellular technologies and open challenges for supporting the data services required. Then we describe network slicing as an enabler to face these challenges. We present an analysis of the cost equilibrium for network slicing to be effective for car manufacturers, and tests in real 5G networks that demonstrate the performance improvement in OTA updates coexisting with other services.

## 1 | Introduction

The fifth generation of mobile network technology (5G) has been conceived to support a wide range of use cases with diverse quality of service (QoS) requirements, including those on data rate, latency, reliability, and mobility, among other key performance indicators (KPIs). 5G standards have introduced new technologies such as network slicing, multi-access edge computing (MEC) and bandwidth parts (BWPs) to tailor network resources to the specific requirements of the services consumed by end users. Additionally, 5G standards incorporate novel features that are particularly relevant for industrial verticals, such as integration with time-sensitive networking (TSN) for the support of future operational technology (OT) industrial use cases requiring wireless connectivity [1], and standalone private networks, offering end users thigh control over aspects such as coverage and data privacy [2]. In the automotive field, 5G includes specific features designed for vehicle-to-everything (V2X) communications [3].

The automotive industry encompasses a wide range of multiply connected devices with very diverse requirements, from factory machinery to vehicles. The data traffic exchanged by these devices have different levels of dependency on latency, data rate, or reliability, as well as different precedence levels. Network slicing is a key technology for mobile networks to support various use cases cost-effectively through a common infrastructure. This technology allows operators to deploy and remove multiple logical networks tailored to the requirements of predefined use cases, unlike the conventional one-size-fits-all approach followed by previous mobile networks. These logical networks are isolated across the core network (CN) and the radio access network (RAN), ensuring that the traffic loads in a network slice do not impact the performance of the traffic in other slices.

This paper is the first thorough assessment of 5G network slicing from an automotive industrial sector perspective. It is organized as follows: Section 2 introduces network slicing, including its networking basics and the requirements of the automotive industry in relation to this paradigm. The state of the art of network slicing in the automotive industry is discussed in Section 3. We comment on recent advances related to industrial and vehicular communications and propose open challenges given the limitations of previous cellular communication systems. Section 4 analyzes the cost equilibrium for the coexistence of automotive and regular traffic in operator networks. Section 5 presents numerical tests on a real 5G SA network to demonstrate the benefits of network slicing for the coexistence of those traffic types. Then, Section 6 describes a real on-the-road test of a vehicle software update based on network slicing. Finally, Section 7 concludes the paper.

## 2 | Context

In this section we provide the context of our study: the requirements of the automotive industry during manufacturing and post-sales support, the basics of 5G architecture and network slicing in particular, and its general benefits.

### 2.1 | Requirements of the Automotive Industry

The communication requirements of the automotive industry can be categorized into factory and post-sales (vehicle) requirements.

It is possible to define numerous factory use cases with highly diverse QoS network requirements. Nowadays, there exist machinery equipped with multiple network interfaces using different technologies to fulfill the specific requirements of the services they consume. Examples of these use cases include remote operation, which requires uninterrupted and reliable connectivity; quality control, involving high-quality video streaming that requires high data rates and low latencies; and wireless sensor network (WSN) communications, which prioritize reliability over delay or data rates [4]. In addition, at any given time, some of these cases may preempt others, depending on aspects such as real-time deadlines or criticality of remote actions, hence the usefulness of network slicing in the combined scenario. Moreover, in many industrial scenarios, downtime and malfunctioning significantly reduce revenue [5].

On the road, the experience of car manufacturers such as Stellantis is that vehicle electronics lifecycles differ significantly from those of domestic consumer electronics. From the initial concept to the start of production (SOP), the development phase spans approximately 3 years. Once in production, a vehicle model typically remains in circulation for 6 to 7 years, depending on market reception and demand. Around the midpoint of this lifecycle, the vehicle often undergoes a major facelift, incorporating updates, new engine models, and new advanced driver assistance systems (ADAS) or other features to meet evolving market expectations and enhance the driving experience. During these pivotal stages, critical networking components are frequently upgraded or replaced.

The communication module of a vehicle, known as the telematic control unit (TCU), is selected by manufacturers based on a variety of factors, including regulatory requirements, market-specific considerations, vehicle trim levels, and strategic objectives. This involves substantial validation efforts to ensure that the quality of the vehicle is not compromised. To streamline complexity and reduce development costs, manufacturers often install selected TCU models in multiple vehicle lines. As a result, any modifications to TCU specifications typically incur in significant expenses and delays.

Today, TCUs support a wide range of use cases including over-the-air (OTA) updates, high-definition map updates, ADAS, in-vehicle entertainment, e-commerce/app stores, Wi-Fi hotspots, OEM telematics, and remote services. However, with the emergence of new network technologies and the rapid evolution of user demands, the services consumed by vehicles, as well as their requirements, are expected to evolve. Consequently, technological solutions must be designed to meet diverse future requirements without replacing the vehicle electronics.

The case of Stellantis is very representative. Stellantis is currently the fourth car manufacturer worldwide in units sold, and the second in Europe after the Wolkswagen group.[†] It comprises brands such as Citroën, Chrysler, Jeep, Alfa Romeo, and Dodge. In [6], Stellantis reported 6 M OTA updates during 2021 for a fleet of 12 million connected vehicles, and a forecasted value of 400 M updates in 2030, for a fleet of 34 million connected cars. In 2024 the overall data exchanged by a major manufacturer and its vehicles will be in the 10 PB range. It should be noted that, from a business model perspective, mobile network operators (MNOs) are expected to introduce new alternatives beyond payment per GB, adequate for background data transfer [7] for OTA updates.

### 2.2 | Network Slicing

5G standards define procedures for the complete management of the lifecycle and operation of network slices. Departing from a basic overview of the architecture of the 5G core network, this subsection provides a comprehensive overview of network slicing, its related procedures and the benefits it brings.

#### 2.2.1 | Architecture of the 5G Core Network

Figure 1 shows the standardized architecture of the 5G core network, defined in [8]. It was specifically designed to support the virtualization of its constituent entities, known as virtual network functions (VNFs). These are software modules that can be executed on standard computing equipment. This modular architecture facilitates lifecycle management by operators, allowing them to dynamically adjust the infrastructure to the ongoing demands of end users. All core network entities are interconnected through service-based interfaces, where VNFs expose their available services as HTTP/2 methods [9]. The network repository function (NRF) collects and catalogs the services exposed by existing VNF instances within the network. VNFs interact with the NRF to obtain information about functions that provide the specific services required to complete standard procedures [10].

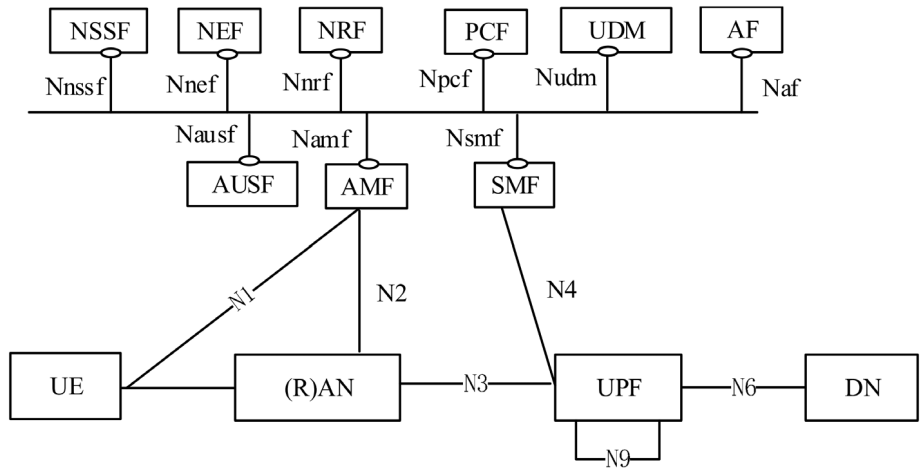

**FIGURE 1** | 5G System architecture [8].

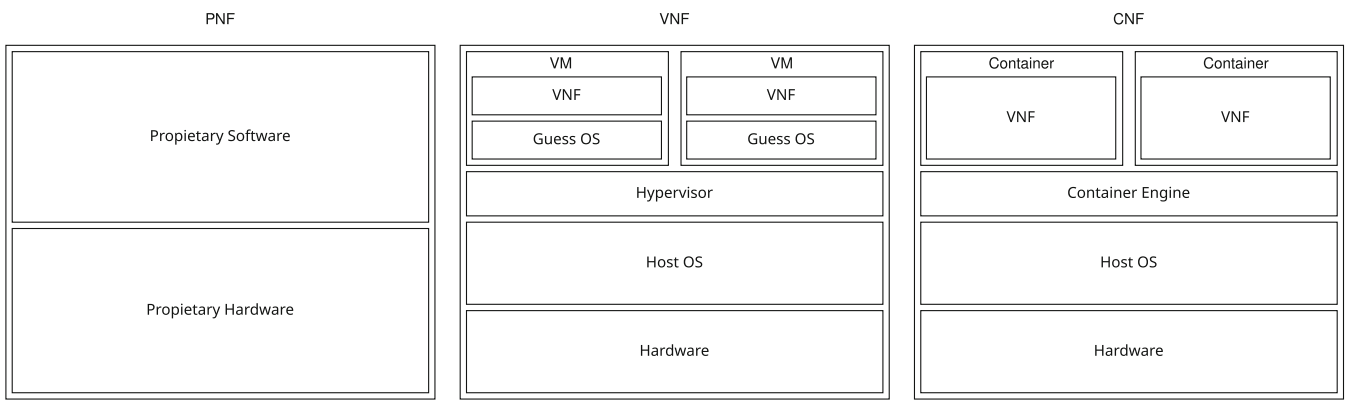

**FIGURE 2** | Different architectures of network functions.

Figure 2 illustrates different approaches for the deployment of network functions. Traditionally, mobile networks had followed the physical network function (PNF) approach, where network entities were deployed as independent proprietary hardware devices. In recent years, the mobile network industry has shifted from this hardware-centric model to a software-centric paradigm, adopting the principles of network function virtualization (NFV). By running VNFs on standard commercial equipment, they can share computing resources, dynamically adjusting their usage in real-time based on current loads. Additionally, surplus computing resources can be leveraged to accommodate third-party services, in line with the MEC paradigm, thereby providing end users with proximity computational services [11].

Two softwarization solutions stand out in the software-centric paradigm: virtualization and containerization. Their main difference is that virtualized VNFs are run in independent virtual machines (VM) with their own kernel, fully isolated from the VMs of other VNFs, while containerized VNFs run in containers, which are lighter elements that share the kernel of the host machine, therefore reducing the required computing load at expense of less isolation among VNFs.

### 2.2.2 | Network Slicing in 5G Networks

A network slice instance consists of a set of VNFs, a series of network interfaces interconnecting them, and a collection of configurations and policies tailored to the requirements of a specific use case. Each network slice is designed to achieve target network performance metrics, measured by KPIs such as peak and average data rates, latency, reliability, and predictability, among others.

The lifecycle of a network slice instance is managed by templates known as network slice templates (NSTs). Each NST specifies the physical and logical resources required to deploy a network slice instance, along with any additional policies to be executed during its deployment, if applicable. This topic is thoroughly addressed in [12], which outlines various high-level solutions for automating tasks such as on-demand slice deployment, reconfiguration, and failure management. Since there is no standardized low-level implementation for these mechanisms, manufacturers must develop proprietary solutions to achieve the high-level objectives.

The operator can deploy multiple independent network slice instances with the same or similar configurations for different reasons, including providing services of the same class across different locations, offering differentiated QoS within the same service type (e.g., priority versus non-priority users), and enhancing network security through increased isolation of control and data flows.

Regarding isolation, various network slicing deployment approaches offer different levels of isolation between VNFs and computing resource efficiency. A single VNF can participate in multiple network slice instances, even based on different network slice templates. Conversely, a VNF can be deployed in an exclusive computing instance for maximum isolation. Intermediate solutions involve VNFs associated to independent slices being executed in shared computing instances. In practice, operators can combine these approaches, providing higher isolation for network slice instances supporting critical use cases, despite the higher computing resource requirements, while less critical network slice instances can adopt more resource-efficient approaches, even if this results in lower availability or performance.

### 2.2.3 | Network Slicing in RAN

Many vehicle use cases will coexist not only in the network but also in the vehicle terminal, so that, in addition to network slicing to support services with diverse requirements, RAN slicing will be also highly useful.

Mobile transmissions are scheduled by base stations, which notify end devices in advance about the time periods and frequencies they should use for transmitting and receiving data frames. The base station allocates channel resources, known as physical radio blocks (PRBs), to end users, based on the characteristics of the corresponding data traffic. Base stations can be configured to guarantee to specific data flows a minimum data rate, reduced delay, or target reliability, by setting the frequency and allocating resources to each traffic, as well as specific queue management algorithms. In this regard, the 3rd Generation Partnership Project (3GPP) has specified in the 5G standards that manufacturers must implement mechanisms in their base stations for differentiated traffic processing based on network slices. However, the implementation of this feature is entirely left to the manufacturers [13].

Complementing PRB scheduling, 5G introduced a new technology called bandwidth parts (BWPs) to support multiple verticals simultaneously. A BWP consists of a subset of PRBs. These subchannels are self-contained, meaning that all control and data transmissions occur within the defined frequency range. This approach enables simple user equipments (UEs), with limited capabilities and power constraints, to operate within a general network. This is in contrast to previous mobile technologies, where control transmissions were spread across the entire channel, so that independent narrow-band carriers were necessary for simple devices such as sensors and actuators to operate.

The channel is dynamically split into BWPs in real-time according to a pattern that the network shares with the end devices. Thus, BWP allocation is not static, but 5G networks are able to distribute channel resources among the use cases with pending traffic based on the precedence and requirements of the different verticals.

### 2.2.4 | Procedures Related to Network Slicing

Network slices are identified within the 5G network by a label known as single-network slice selection assistance information (S-NSSAI). This identifier is included in all messages related to network slicing to refer to the specific network slice to which the requested action applies. The S-NSSAI consists of two fields: the slice/service type (SST) and the slice differentiator (SD). The SST indicates the traffic class (e.g., enhanced mobile broadband (eMBB), ultra-reliable low latency communications (URLLC), etc.) that the network slice is tailored to, while the SD is used to differentiate levels of treatment (e.g., premium vs. best effort) within the same traffic class. The list of S-NSSAIs that the end user is permitted to request is stored in the unified data management (UDM), along with the allowed data network names (DNNs).

Figure 3a illustrates the registration procedure in 5G networks. During registration, the UE can preemptively indicate the list of S-NSSAIs it expects to use in the upcoming session. The network then verifies that the subscriber is authorized to operate in these network slices, confirms that the base station to which the end device is connected supports all the requested network slices, and configures the necessary VNFs if required.

To operate in a network slice, the UE must request the establishment of a packet data unit (PDU) session using the corresponding S-NSSAI label. The PDU session establishment procedure is depicted in Figure 3b. Each PDU session is managed independently, configured in the end device as a separate logical interface, and assigned a unique IP address. As a result of this procedure, a new tunnel is established between the UE and a user plane function (UPF) to forward the traffic through the network slice.

The mobile network configures end devices with routing tables to map data flows to network slices using the UE route selection policy (URSP) mechanism. For each subscriber, the network determines a set of routing rules based on traffic filters that point to specific S-NSSAIs and DNNs, which together form a URSP table. Traffic filters can focus on parameters such as IP addresses, transport protocols, ports, and application IDs. Each traffic filter is associated with one or more route selection descriptors (RSDs), which indicate, in order of precedence, the most suitable S-NSSAI, DNN, radio access technology (RAT), and/or session and service continuity (SSC) mode for the corresponding traffic. The URSP table is typically configured on the UEs after registration but can be updated at any time to adapt to the dynamic conditions of the network and the QoS requirements agreed upon with the subscriber.

For each new data flow, the UE evaluates its URSP table to determine the most suitable S-NSSAI. The UE then checks if a PDU session has already been established for that S-NSSAI. If no such session exists, the UE attempts to establish a new PDU session. If the PDU session cannot be established—due to the network being unable to deploy the required VNFs, for example—the UE repeats the procedure with the next RSD associated with the matched traffic filter. Once a suitable PDU session is established, a new forwarding rule is set in the operating system of the UE to route the traffic of the corresponding data flow through the logical interface associated with the relevant network slice.

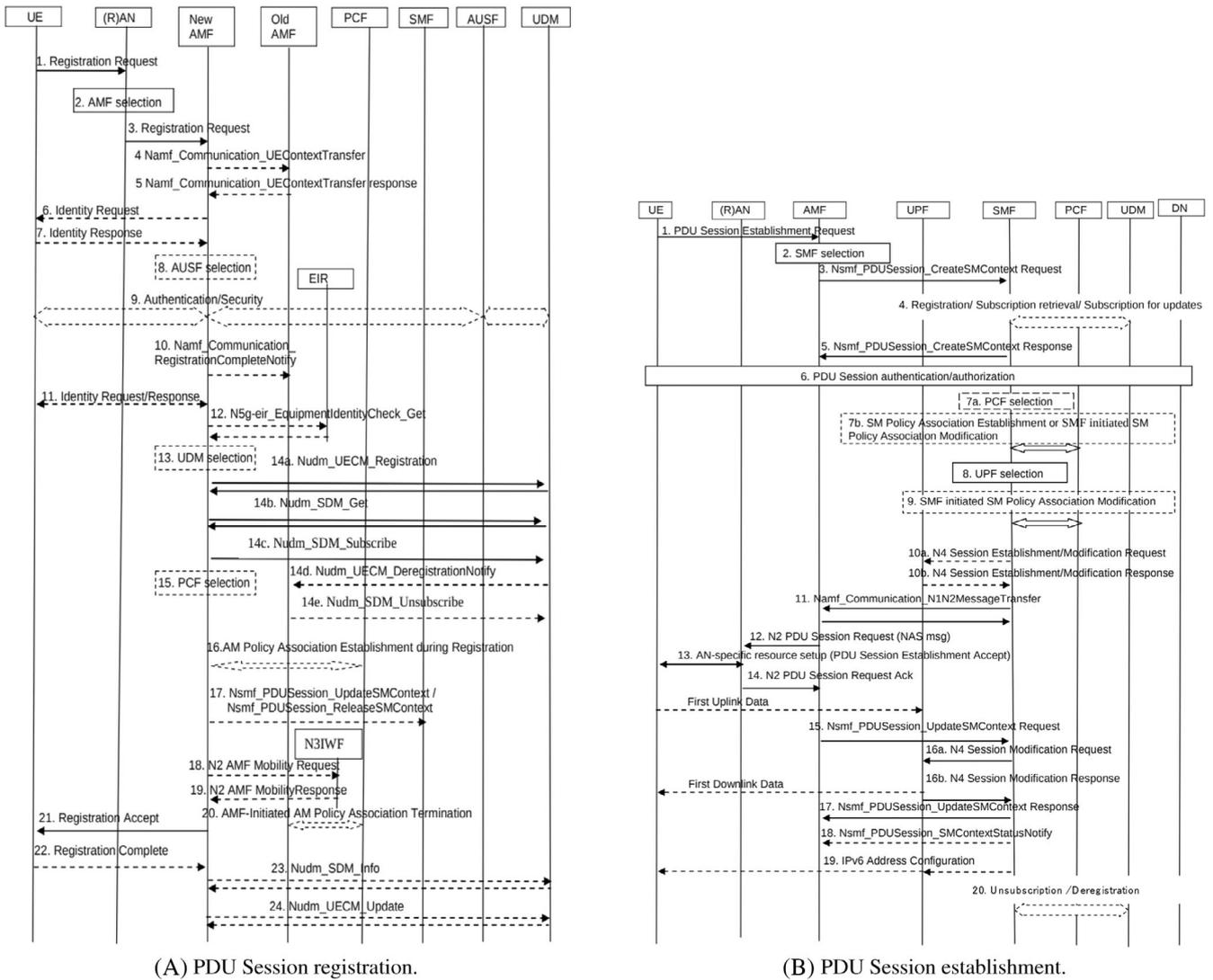

**FIGURE 3** | PDU Session registration and establishment [14].

### 2.2.5 | Key Benefits of Network Slicing

By implementing multiple independent logical networks on top of a shared physical infrastructure, mobile operators can more effectively meet the diverse requirements of the traffic they need to accommodate.

The primary benefit of network slicing for data flows is enhanced security and isolation. By isolating network nodes, mobile networks can ensure reliable operation and performance, irrespective of the loads carried by other network slices. Additionally, a malfunction in a VNF only affects the performance of the equipment connected to the specific network slice instance, thereby increasing overall network availability. This isolation also provides greater protection against data breaches, securing sensitive data in critical use cases such as surveillance, e-health, and energy plants [15, 16].

Another key benefit of network slicing pertains to QoS. Network slicing enables operators to define coarse-grained traffic processing rules that meet the general requirements of different use cases. By applying specific configurations and policies to each network slice instance, operators can ensure compliance with KPIs such as data rate, latency, and reliability. For instance, use cases with stringent latency requirements but less dependence on reliability can be processed through a network slice that provides shorter resource allocation times and stricter queue management algorithms, albeit with reduced data rates compared to a typical eMBB network slice [17].

In this regard, it is important to note that QoS can be further enhanced through the use of QoS flows, which are designed to handle specific data flows and specify special traffic processing treatment to be implemented by forwarding nodes along their data path. However, this process requires activation per user's data flow via an exposure application programming interface (API) as a request from either the server-side or client-side application. Therefore, network slicing is envisioned as the primary solution for managing QoS in standard use cases that do

5 of 13

not require specialized behavior, aligned with the KPI templates established by operators to formulate NSTs.

Finally, network slicing enhances network scalability and efficiency by enabling shared infrastructure resources across different verticals supported by the network. This capability allows for dynamic resource allocation to network slice instances based on real-time traffic loads, prioritizing different use cases as needed to optimize network performance in a cost-effective manner [18].

## 3 | State of the Art and Related Work

This section explores the necessity of 5G communications in factories and the ways in which 5G technology enhances vehicle communications, with a particular emphasis on network slicing. From the review of the state of the art, we offer a comprehensive overview of the most pertinent open challenges that must be addressed.

Network slicing, as a general-purpose technology, has become a prominent topic in the literature. The paradigm is comprehensively explained in [19] and [20] The authors of [19] outline the requirements and challenges for its implementation. More in-depth issues, such as RAN virtualization, holistic and intelligent slice orchestration, and secure sliced networks, are explored in [20]. In [21], the authors survey research on network slicing across three layers, infrastructure, network function and service, as well as management and orchestration (MANO). Additionally [22], provides a high-level overview of network slicing, discusses the organizations involved in its standardization, and identifies open issues such as end-to-end slicing, network slicing in roaming scenarios, and the migration and interoperability of VNFs. Admission control to network slices via probabilistic forecasting is studied in [23, 24].

However, the majority of prior work predominantly addresses the needs of operators and vendors in implementing the paradigm within general-purpose 5G networks. There is a noticeable gap in the literature regarding the application of network slicing from the perspective of industrial verticals such as automotive use cases and their specific requirements. This section explores how implementing network slicing can enhance efficiency, functionality, quality, scalability, and other KPIs in the automotive sector. We examine its benefits both during the vehicle manufacturing process and throughout its operational life on the road.

### 3.1 | Network Slicing in the Automotive Factory

Prior to 5G, mobile networks were primarily designed as general-purpose connectivity platforms with limited capabilities to differentiate across various use cases. In contrast, 5G technology enables the creation of an ecosystem for technical and business innovation, encompassing vertical markets such as automotive and manufacturing. This has garnered significant attention, as evidenced by the substantial body of research discussing how 5G networks can facilitate the advancement of industrial processes [25–28].

The requirements of Industry 4.0 and the technologies driving its implementation, along with the use cases enabled by 5G, are outlined in [25]. In [26], the authors review the trends and challenges across various industries such as IoT, automotive, manufacturing, smart factories, smart grids, and smart cities. They highlight the key enabling technologies and capabilities introduced by 5G to facilitate digital transformation. Several Industry 4.0-related tactile Internet applications supported by 5G, owing to its enhanced performance over previous generations, are discussed in [27], along with their network communication requirements. The benefits of 5G for manufacturing processes in smart factory environments are detailed in [28], which also outlines the industrial challenges involved. In particular, regarding the benefits of network slicing in factory processes [4], offers a high-level description of the creation of logically separated, use-case-specific virtual networks on top of a general physical network. The focus is on network slice management, potential strategies, and implications for a smart factory. The analysis in this work can assist network infrastructure managers in factories to choose the most reasonable strategy for tailoring a network to specific communication requirements.

### 3.2 | Network Slicing for Vehicle Communications

Vehicular communications represent one of the most challenging use cases for 5G networks. This vertical comprises the provision of standard communication services, such as data connectivity to cloud applications on the Internet, as well as communications specifically related to vehicular use cases, such as safety-related V2X communications.

V2X specifications were first introduced in mobile standards in 3GPP Release 14 [29] as an evolution of the ProSe (Proximity Services) in releases 12 and 13. 3GPP has outlined a road map for defining C-V2X functionalities, divided into three well-defined phases. The initial phase, based on 4G, primarily focused on facilitating V2V communication through device-to-device (D2D) communication and basic security services based on vehicular-to-infrastructure (V2I) communication [30]. The second phase enhanced existing LTE standards, introducing advanced features and new use cases such as vehicle platooning, extended sensors, advanced driving, and remote driving [31]. At this stage, the first version of V2X standards based on the 5G NR standards, 5G-V2X, was introduced in Release 15. In the third phase, starting with Release 16, the focus shifted to providing advanced features and performance improvements based on the new iterations of the 5G NR standard [30].

The impact of network slicing in 5G V2X communications has been studied in the literature. Most existing works explore the capabilities and requirements of this technology, while some provide validation and results based on different simulations. Despite the clear interest in this topic, there is a notable lack of research evaluating real-world implementations.

The survey in [32] explores the challenges and requirements for V2X use cases, providing a historical overview of major milestones, standards, and enabling technologies, including network slicing, with a focus on vehicular communications. In [33], the authors propose an intelligent network-slicing architecture to enhance QoS in vehicle-to-vehicle communications, validated through simulation. In [34], the authors emphasize the importance of network slicing for enabling vehicular edge computing and ensuring the required QoS for reliable V2X communications.

The authors of [35] discuss the challenges of network slicing security in 5G-V2X scenarios, presenting a schema based on federated collaborative learning for detecting inter-slice attacks with high accuracy using a small model size. In [36], the authors propose a strategy to coordinate different V2X operating modes and introduce a novel algorithm, latency-aware mode coordination (LAMOC), to reduce network latency in vehicular network slicing.

### 3.3 | Open Challenges of Network Slicing in the Vehicular Environment

The commercial connected vehicle era began in 1996 with the introduction of communication capabilities in vehicles, initially aimed at providing emergency communications and enhanced security services [37]. Since then, connected vehicles have evolved into a comprehensive enabler for numerous in-vehicle mobility and non-mobility use cases and services. These include in-vehicle entertainment, e-commerce and app stores, high-definition maps, over-the-air (OTA) software updates, Wi-Fi hotspots for connecting mobile phones and other end-user devices, ADAS, OEM telematics, and remote services.

Significant differences exist between these use cases in terms of network requirements and priority. In 4G, these differences were managed by using different access point names (APNs), which involve distinct gateways between the mobile network and the service-providing network. APNs facilitate the separation of diverse data flows and remain the standard solution of MNOs for split billing and traffic routing purposes.

Given that the cost of networking services is directly related to the volume of data exchanged, the capability to charge based on the service or type of traffic is crucial for developing viable business models for connected services and managing vehicle connectivity expenses. 5G has introduced two new VNFs related to billing: the charging function (CHF) and the charging enablement function (CEF) [38]. The CEF collects and process information related to data usage, including the amount of traffic transmitted or received by network slice for each subscriber. The CEF communicates with the CHF to charge users according to their communications [39]. However, this functionality is not implemented today by most MNOs, specially because it conflicts with the traditional strategy of charging on the basis of the subscription contracted by the end-user, and therefore requires new and more complex marketing strategies.

With regard to traffic routing, MNOs typically offer two types of APNs: private and public. Private APNs are used to route traffic to the cloud of the OEM, typically for telematics and remote supporting services, but they can also support some third-party services. Public APNs, on the other hand, provide direct access to the Internet without forwarding traffic through the backend of the network of the OEM. Additionally, with the advent of MEC, some MNOs are proposing specific APNs for local routing.

Furthermore, regulation may also impact the APNs being used for vehicular communications. Electronic communications are subject to diverse regulatory frameworks across different countries, and service separation through multiple APNs can be a viable solution to meet these varying mandatory requirements.

However, there are stringent limitations on the number of APNs that can be used by end devices in mobile networks. In LTE, the standards impose a limit of 8 APNs that a terminal can maintain simultaneously, with many LTE modems supporting simultaneous operation of up to 7 APNs only. This limitation is critical when segregating data flows by communication needs in automotive environments, encompassing both vehicle communications and manufacturing processes.

Finally, despite its technical feasibility, operators are not offering APNs as a means of providing QoS. Even though APNs may use different configurations for differentiated traffic processing, operators do not guarantee specific network performance KPIs for each gateway (i.e, low latency, high throughput, best effort...). This is mainly due to their current business model for connectivity, which is mainly based on a price per gigabyte.

5G network slicing can effectively address those limitations, as it enables the isolation of data flows across different traffic categories. It enhances reliability, facilitates differentiated billing support, and simplifies routing configuration by allowing the network to configure routing rules within end devices, determining the appropriate network segment for each data flow since its origin.

However, some issues still must be addressed. 3GPP has standardized SST values for six specific traffic classes, shown in Table 1, to facilitate interoperability among mobile operators [8]. These SSTs must be universally recognized by all 5G networks, enabling devices to request data session establishment within network slices designed for these traffic classes, even during roaming. The predefined set of traffic classes does not fully accommodate the diverse requirements of communications in vehicles, which often are connected through roaming due to their mobility. Moreover, in the case of roaming, the interpretation of the SD field can vary among operators. For example, given a specific SST, operator A may interpret SD = 0 as best-effort traffic, while operator B as might define it for premium communications.

**TABLE 1** | Standardized SST values.

| Slice/Service type | SST value | Characteristics |
| --- | --- | --- |
| eMBB | 1 | Slice suitable for the handling of 5G enhanced Mobile Broadband. |
| URLLC | 2 | Slice suitable for the handling of ultra-reliable low latency communications. |
| MIoT | 3 | Slice suitable for the handling of massive IoT. |
| V2X | 4 | Slice suitable for the handling of V2X services. |
| HMTC | 5 | Slice suitable for the handling of High-Performance Machine-Type Communications. |
| HDLLC | 6 | Slice suitable for the handling of High Data rate and Low Latency Communications. |

Another critical aspect concerning QoS in network slicing is the current inability to request the establishment of a PDU session based on target network performance metrics such as specific data rates, latency, and reliability. Currently, the standards only support end devices requesting S-NSSAIs aligned with the RSDs corresponding to their respective data flows. The allowed set of S-NSSAIs for each subscriber, along with their traffic flow mapping, is typically configured at the moment the subscriber acquires the communication services. There is a lack of standardized procedures for modifying this configuration, making it static in practice. To address this challenge, the GSM Association (GSMA) has defined an endpoint allowing end users to request temporary activation of an enhanced network slice for specific application traffic, operating under a pay-per-use model [40]. Furthermore, individual data flows can request specific QoS behaviors from operators through client- or server-side applications, although this feature may not be universally available across all operators for all subscribers.

Finally, while the number of concurrent data sessions was increased to 16 in 5G standards, this may be insufficient and still represents a limitation given the broad diverse range of applications expected for implementation in the near future. It will be necessary to harness the strong influence by network slicing implementations for consumer electronics, given the widespread adoption of Android in car dashboards. Smartphones, and particularly Android OS, is leading the way with support for network slicing already available in Android 12 [41], which is even proposing slicing at application level. The development solutions of car manufacturers for third parties, such as [42] will likely drive entertainment-oriented applications for user terminals connected via on-board Wi-Fi. Given the coexistence of these user applications with the growing needs of OTA traffic, as mentioned in Section 2.1, in addition to the technical challenges discussed, there will be a clear need to manage different types of traffic with very different requirements at the vehicle, and network slicing seems an adequate solution for this challenge.

## 4 | Coexistence of Automotive and Regular Traffic: Cost Equilibrium Analysis

One of the principles of Industry 4.0 is to enhance efficiency by improving manufacturing processes to produce more goods while reducing operational expenses (OPEX). In terms of networking communications in vehicular and factory environments, as aforementioned, there exist today many traffic categories with very different priorities, transmission sizes, and networking requirements. Today, MNOs offer a very limited range of subscription options to end users to receive traffic processing tailored to user requirements. Existing subscriber plans are based on different pre-defined amounts of mobile data with fixed monthly costs, combined with a variable charge depending on the data used above the plan. Regardless of the subscription types, traffic from different users is generally treated with the same priority, although the users with more expensive plans, and therefore with higher data usage limits tend to make greater usage of the network. In fact, nowadays, MNOs do not offer any means of charging differentiation for the different traffic classes that a single device may handle.

Although automotive traffic categories such as emergency calls, critical manufacturing procedures, critical OTA updates and entertainment may require traffic processing and enhanced treatment, other communications such as cartographic and other non-priority updates do not need to be completed faster. Therefore, automotive manufacturers can benefit from differentiated charging based on traffic categories, with reduced costs for transmissions that only receive sporadic resources. However, with such an approach, low-priority traffic may starve.

For a MNO, guaranteeing SLAs during intervals of high network loads is of the utmost importance. Assuming that the user load during high-load periods is directly related to the data limits of the subscription plan of end users, and that the cost of mobile network subscriptions is based on the expected data usage during high-load periods, the cost of a type $i$ mobile network subscription can be expressed today as:

$$c_i = k + x_i \cdot \frac{\alpha_i \cdot pl_{tx}}{t_{tx}}, \tag{1}$$

The cost function $c_i$ (1) has a fixed cost parameter $k$ and a variable term $x_i$ that depends on the amount of data transmitted during high-load periods, $\alpha_i \cdot pl_{tx}$, where $\alpha_i$ represents the average difference on data usage compared to the average of subscribers with a mobile subscription of type $i$, with $\sum_i \alpha_i = 1$ and $\alpha_i > 0$, $pl_{tx}$, and $pl_{tx}$ is the average transmission size during a high-load period; and $t_{tx}$ is the average time to transmit $pl_{tx}$ bytes, is the same for any user, representing the current scenario where the aggregated traffic from a subscriber has the same priority of that perceived by the rest of the subscribers.

The cost function of a network that implements resource allocation to the overall traffic according to traffic categories without seeking priority fairness among users can be defined as:

$$c'_i = \sum_j c'_{i,j}, \tag{2}$$

where the cost per subscription type $i$ and traffic category $j$, $c'_{i,j}$, can be expressed as:

$$c'_{i,j} = k' + x'_i \cdot \frac{\beta_j \cdot \alpha_i \cdot pl_{tx}}{t_{tx}}, \tag{3}$$

with $k'$ and $x'_i$ being the fixed and variable terms of the cost function, and $\beta_j \in [0, 1]$ representing the average proportion of network resources allocated to traffic category $j$ during high-load periods, with $\sum_j \beta_j = 1$.

A background traffic category $j = 0$, with zero priority over other communications, would not receive network resources during saturation, and therefore $\beta_j = 0$, leading to $c'_{i,0} = k'$. The resources that are not allocated to traffic category $j$ with this resource allocation strategy can be used to accommodate more users during high-load periods or/and implement new traffic categories with enhanced performance at higher costs.

Let $N_i$ and $N'_i$ represent the average number of subscribers transmitting during high-load periods for the resource allocation strategies related to Equations (1) and (2) per subscription type $i$. In order for the operators to retain the same profit, then:

$$\sum_i c_i = \sum_i N_i \left( k + x_i \cdot \frac{\alpha_i \cdot pl_{tx}}{t_{tx}} \right) = \sum_i c'_i$$
$$= \sum_i N'_i \left( k' + \frac{\sum_j x'_i \cdot \beta_j \cdot \alpha_i \cdot pl_{tx}}{t_{tx}} \right), \quad (4)$$

which results in:

$$k' = \frac{N \cdot k + \sum_j \left( \alpha_i \cdot \left( N_i \cdot x_i - \sum_j N'_i \cdot x'_i \cdot \beta_j \right) \right)}{N'} \cdot \frac{pl_{tx}}{t_{tx}}. \quad (5)$$

In order for the traffic category $j = 0$ to be charged less when the operator implements the proposed resource allocation approach:

$$c'_{i,0} = k' < c_i = k + x_i \cdot \frac{\alpha_i \cdot pl_{tx}}{t_{tx}}. \quad (6)$$

To meet equality in Equation (4), then:

$$\sum_i N_i \left( k + x_i \cdot \frac{\alpha_i \cdot pl_{tx}}{t_{tx}} \right)$$
$$< \sum_i N'_i \left( k + x_i \cdot \frac{\alpha_i \cdot pl_{tx}}{t_{tx}} + \frac{\sum_j x'_i \cdot \beta_j \cdot \alpha_i \cdot pl_{tx}}{t_{tx}} \right), \quad (7)$$

which can be expressed as:

$$(N - N') \cdot k < \sum_i N'_i \left( \sum_j x'_i \cdot \beta_j \cdot \alpha_i \right) - \sum_i (N_i - N'_i)(x_i \cdot \alpha_i), \quad (8)$$

which is met if $N' \geq N$ or if the variable term of the cost per traffic category, $x'_i$, is increased for regular and high-priority communications. Naturally, this extra cost can be disseminated only between high-priority traffic classes, so regular communications can preserve the original charge rates.

Given the fact that large-scale car manufacturers will have a strong negotiation capability with MNOs on relatively equal terms, this approach can be used to dimension MNOs' tariffs, given their populations of regular users, the number of cars they are paid to serve, and the services they will require as presented at the beginning of this section.

## 5 | Coexistence of Automotive and Regular Traffic: Traffic Performance Evaluation

In order to demonstrate how network slicing can be a valuable tool to solve the problem, we evaluated the performance of coexisting high-priority and background traffic categories when network resources are allocated based on the traffic class.

We conducted this evaluation in a state-of-the-art 5G Standalone (SA) network, composed of an industrial-grade gNodeB and an open-source core network implemented using the code of project [43]. The network operates in a 50 MHz licensed band leased by a MNO in exclusive access for this research.

The architecture of the 5G network is shown in Figure 4. Two independent network slices were implemented in the core network for high-priority and background traffic, respectively, where the former may represent in-vehicle entertainment and the latter low-priority OTA updates. The gNodeB was configured to allocate network resources first to the high-priority network slice and then allocate the remainder of the frequencies to the background network slice. That is, in case of saturation, all the PRBs of the channel are allocated to the high-priority traffic, while background traffic can be allocated all the channel resources if no high-priority traffic is buffered for transmission. For simplicity, the channel resources within a network slice were distributed proportionally among the subscribers.

The evaluation was conducted using 125 UEs, each composed of a 5G SA modem and a Radxa ROCK Pi 4 SE, an embedded board by a dual ARM A72 core @1.5GHz + quad ARM A53 core @1Ghz processor and 4GB of RAM.[‡] The UEs were implemented using different 5G SA modems: 40 Quectel RM510Q-GLHA, 35 Quectel RM520N-GL, 25 Quectel RM502Q-AE, 21 Quectel RM502Q-GL and 4 SIMCom SIM8300G-M2.

The evaluation sought to compare how traffic was allocated to the high-priority or background network slices in high-load scenarios, by comparing the network resource allocation strategy based on equal sharing among subscribers with an allocation strategy differentiating traffic categories. In the first scenario, all the UEs were configured to use the background network slice, so resources are shared equally, whereas in the second one of the UEs was configured to use the high-priority slice to show how the resources are distributed in this scenario with few priority communications.

In the first scenario, the communications of all the UEs take place in the background network slice. Therefore, network resources are allocated to end users proportionally. A test UE was selected for data rate evaluation. It was configured to begin transmitting 500 MB 30 s after the begin of the evaluation, when the network was in a steady saturation state. The rest of the UEs where configured to transmit since the beginning of the evaluation to saturate the network. In the second scenario, the same methodology was applied, but the test UE transmitted through the high-priority network slice. Therefore, the network was expected to allocate all the PRBs of the channel to the test UE during this transmission, and then reallocate them back to the rest of the UEs proportionally. In both scenarios, the time required to complete the transmission of the test UE, as well as the data rate of all UEs, were measured.

The well-known *iperf3* networking tool was used for traffic generation. The UEs were configured to transmit user datagram protocol (UDP) traffic. The test UE was configured to transmit at 500 Mbps, and the other UEs at 4 Mbps, so that their aggregate requested data rate was also 500 Mbps. This exceeded the maximum data rate that the 5G modems used could obtain from

9 of 13

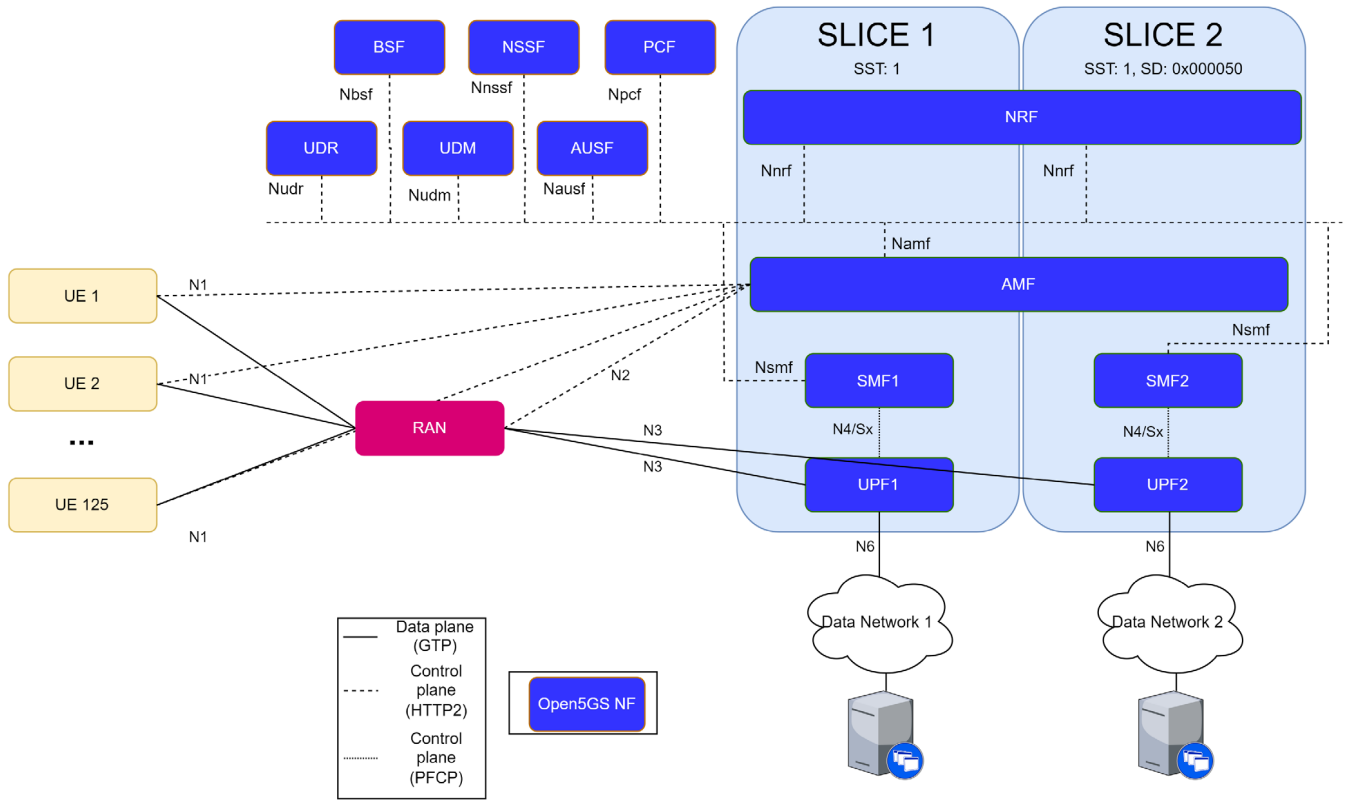

**FIGURE 4** | Architecture of the 5G SA network where the evaluation was conducted.

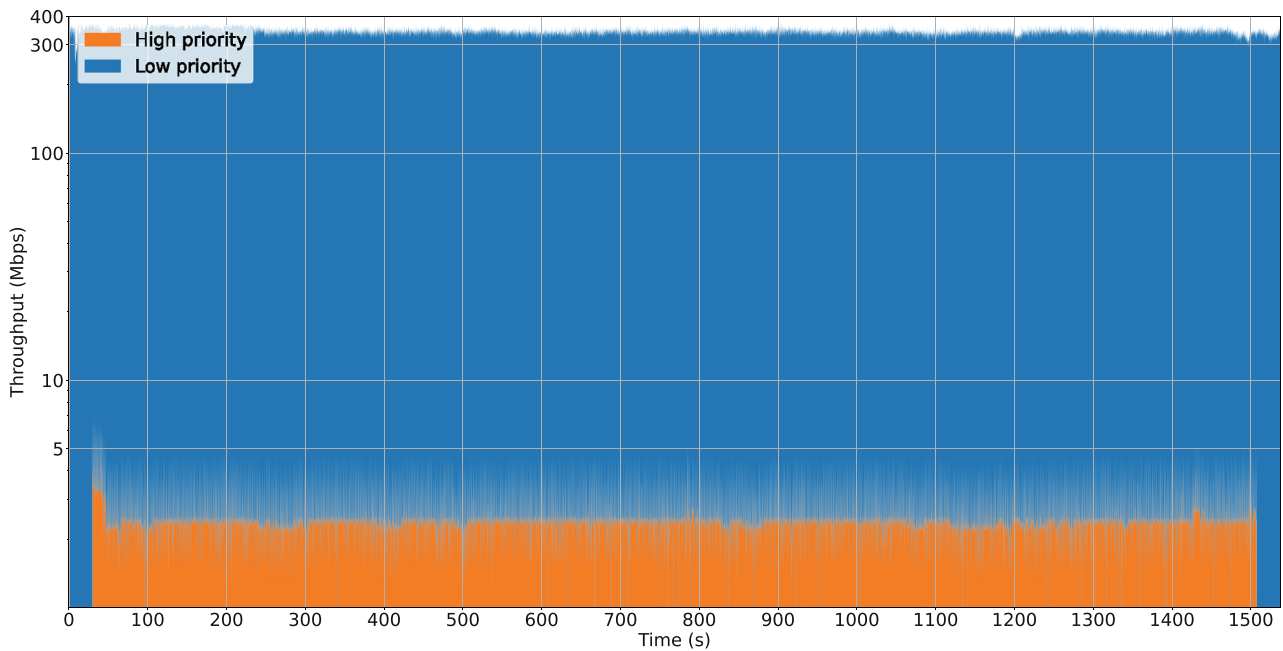

**FIGURE 5** | Data rate, in Mbps, of test and background traffic, for network resources allocated on per-user basis.

the network, for a 50 MHz channel and the modulation schemes supported by the devices, ensuring saturation. The *iperf3* servers were deployed in 10 VMs in a cloud infrastructure directly connected to the 5G core network through a 10 Gbps link.

Figure 5 shows the data rate of the test and background communications in scenario 1 when the end users had equal access priority to the network resources, that is, the approach currently followed by MNOs. In this case, the transmission of 500 MB from the test UE was completed in 1477.41 s, with an average data rate of 2.71 Mbps, while the background traffic was transmitted during the same interval at an aggregate data rate of 337.18 Mbps.

Next, Figure 6 shows the evaluation of scenario 2 when resources are allocated based on the traffic category, with the test communication being transmitted through a high priority slice and background communications through a background slice, which is only allocated remaining network resources not requested by high priority communications. In this case, nearly all channel resources were allocated to the test communication until the transmission of the 500 MB burst was completed, with background traffic receiving residual bandwidth. The test communication was completed in this case just in 12.17 s, at a 327.157 Mbps rate, while the background traffic was transmitted at an average aggregated data rate of 10.37 Mbps.

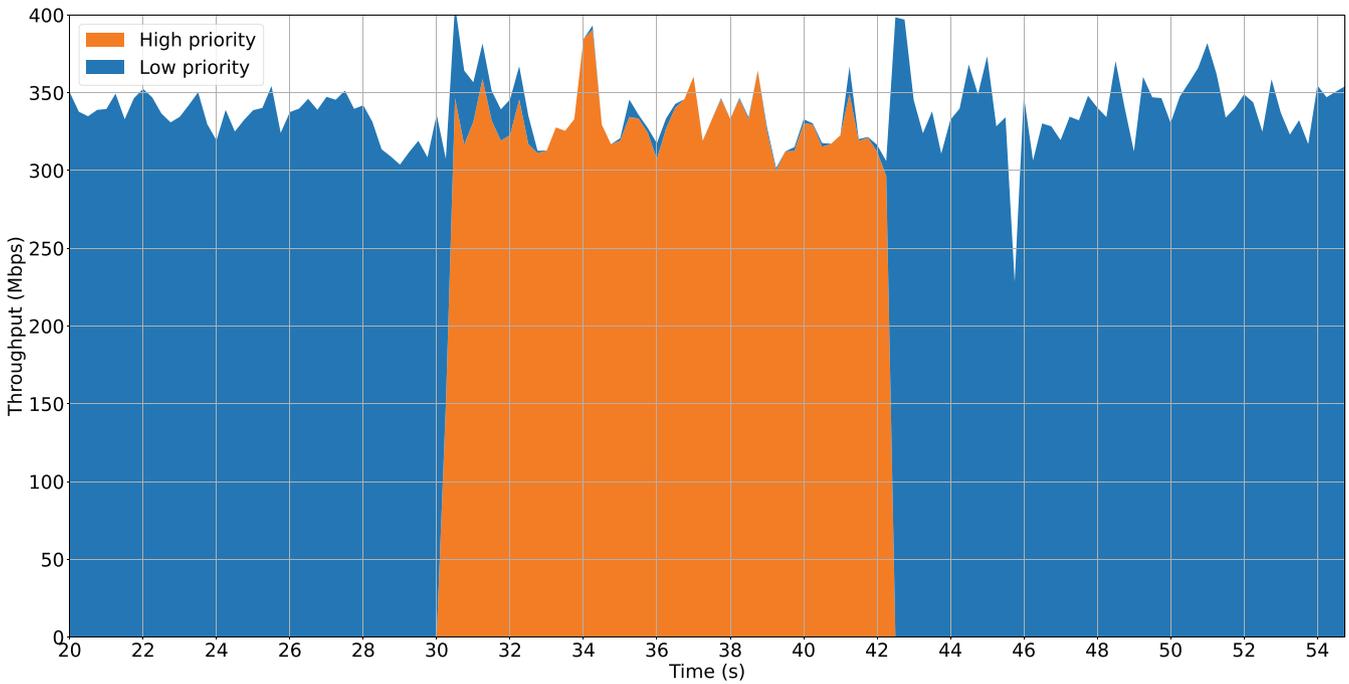

**FIGURE 6** | Data rate, in Mbps, of test and background traffic, for network resources allocated on per-traffic-category basis.

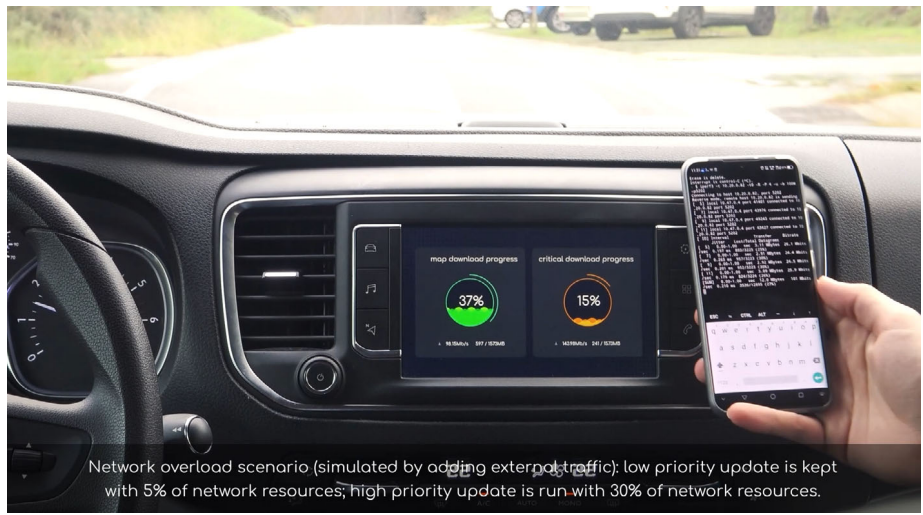

**FIGURE 7** | Snapshot of the vehicle display during the on-the-road test.

For an automotive manufacturer, a non-priority upgrade with no deadline constraints is today transmitted as soon as it has been scheduled, regardless of existing network conditions, as the manufacturer has no incentive in terms of transmission cost to delay the transmission. As shown in Figure 5, this can result in significant performance impairment for the services consumed by third-party end users operating on the area. In the same manner, critical services being consumed by the manufacturer could be severely affected by an external high load, even if the manufacturer is willing to incur in additional cost to meet the requirements of its customers.

On the other hand, a network resource allocation approach based on traffic categories can meliorate higher priority communications, assigning all network resources required to existing high-priority and regular communications and allocating just the remaining resources to background traffic. As a result, the performance of high priority communications is not affected by emerging background traffic, and therefore these communications can be charged with lower prices.

11 of 13

## 6 | On-the-Road Test of Critical Software Update

As a final practical test, two software updates were dispatched to a vehicle connected to the network, one with high priority and the other expecting best effort, emulating the expected behavior for critical and routine software updates. The connection between the network and the vehicle was set with a single Quectel RM500Q-AE modem, with access to two configured slices. One of these slices guaranteed minimum resources, while the second one, serving as the default slice, operated at best effort configuration. The network was deliberately saturated through the best-effort slice, simulating a scenario without free resources available.

When the critical update is requested, the vehicle initiates the download through the priority slice. The network guarantees the download rate for the critical update by allocating dedicated resources to this connection. Given that the network is saturated and there are no free resources available, the network redistributes resources from the best effort slice (thus delaying the routine update). Figure 7 presents a snapshot of the interface displaying this update on the vehicle's display during the test.

## 7 | Conclusions

This paper studies the impact of network slicing on the automotive industry. We have discussed the requirements of the automotive sector for mobile data communications, by considering both manufacturing and post-sales support, and how network slicing enables mobile networks to meet its needs. We have explained that previous mobile communications standards could not satisfy these needs due to technological limitations, and, as a consequence, the challenges that were open at the moment. Then, we have described how 5G network slicing can act as a technological enabler to face these challenges. Next we have studied, both at theoretical and practical level on a real carrier-grade 5G SA network, the coexistence of different services for the automotive sector thanks to network slicing, by considering both cost and performance.

The results highlight how such an approach can help to improve performance of high-priority communications and therefore meet the requirements of critical traffic, and how MNOs can reach a cost equilibrium with car manufacturers for the latter to benefit from reduced tariffs of background traffic.

As future work, we plan to apply artificial intelligence to the management of scenarios involving low-priority massive OTA updates and sporadic latency-critical high-priority OTA updates.

**Acknowledgments**

This work was supported by Xunta de Galicia (Spain), ED431C 2022/04, ED481B-2022-019, IN854A 2020/01, IN854A 2023/01. Ministry of Science, Innovation and Universities (MICIU), State Investigation Agency (AEI), Spain, and the European Social Fund (FSE+), PDC2021-121335-C21, PID2020-116329GB-C21, PRE2021-098290, Universidade de Vigo/CISUG for open access.

**Conflicts of Interest**

The authors declare no conflicts of interest.

**Data Availability Statement**

The data that support the findings of this study are available from the corresponding author upon reasonable request.

**Endnotes**

† Available at https://www.focus2move.com/world-car-group-ranking/, 2024.

‡ Radxa ROCK 4 SE: https://radxa.com/products/rock4/4se/.